\documentclass[conference]{IEEEtran}
\IEEEoverridecommandlockouts
\usepackage{cite}
\usepackage{amsmath,amssymb,amsfonts}
\usepackage{algorithmic}
\usepackage{graphicx}
\usepackage{textcomp}
\usepackage{xcolor}
\usepackage{verbatim}
\usepackage{float}
\usepackage{multirow}
\def\BibTeX{{\rm B\kern-.05em{\sc i\kern-.025em b}\kern-.08em
    T\kern-.1667em\lower.7ex\hbox{E}\kern-.125emX}}
\begin{document}

\title{Comparison of artificial neural network adaptive control techniques for a nonlinear system with delay}

\author{\IEEEauthorblockN{Bartłomiej Guś}
\IEEEauthorblockA{\textit{Institute of Automatic Control \& Robotics} \\
\textit{Faculty of Mechatronics, Warsaw University of Technology}\\
Warsaw, Poland \\
bartlomiej.gus.stud@pw.edu.pl}
\and
\IEEEauthorblockN{Jakub Możaryn}
\IEEEauthorblockA{\textit{Institute of Automatic Control \& Robotics} \\
\textit{Faculty of Mechatronics, Warsaw University of Technology}\\
Warsaw, Poland \\
jakub.mozaryn@pw.edu.pl}
}

\maketitle

\begin{abstract}
This research paper compares two neural-network-based adaptive controllers, namely the Hybrid Deep Learning Neural Network Controller (HDLNNC) and the Adaptive Model Predictive Control with Nonlinear Prediction and Linearization along the Predicted Trajectory (AMPC-NPLPT), for controlling a nonlinear object with delay. Specifically, the study investigates the effect of delay on the accuracy of the two controllers. The experimental results demonstrate that the AMPC-NPLPT approach outperforms HDLNNC regarding control accuracy for the given nonlinear object control problem.
\end{abstract}
\section{Introduction}
The PID control algorithm is a commonly utilized controller in industrial applications. This wide utilization can be attributed to the limited number of adjustable parameters, including the proportional, integral, and derivative actions, simplifying the controller's operation. However, for nonlinear systems, accurate control using constant parameter values for the PID control algorithm is only possible at specific operating points.
\par The limitations as mentioned above of the PID controller have driven the search for alternative control techniques that can ensure satisfactory performance over a larger operating range, especially for nonlinear objects with changing conditions. Various neural-network-based adaptive approaches have been proposed in the literature as promising solutions for controlling nonlinear systems.
%
\par The Hybrid Deep Learning Neural Network Controller (HDLNNC) was proposed in \cite{b2} \cite{b3}. This controller utilizes multiple types of artificial neural networks (ANN), including the self-organizing Kohonen map (SOM), Hebbian learning procedure \cite{b4}, and adaptive learning rate derived from the direct Lyapunov method, to achieve precise control for nonlinear objects, even in the presence of time-varying parameters. An extension of this approach involves considering changes in the dynamics of the controlled object to enable a smooth transition of the control signal.
%
%
\par Another adaptive control strategy gaining popularity in industry is based on the Model Predictive Control (MPC) algorithm, which has been recently extended using nonlinear models, including those in the form of ANN, as reported in \cite{b7} 
\cite{b9}. Nonetheless, employing a nonlinear model directly results in a nonlinear quadratic optimization problem, necessitating more computational resources than a linear model. In \cite{b7}, various approaches for efficient Model Predictive Control with nonlinear models have been proposed, among which the ANN-based Model Predictive Control with Nonlinear Prediction and Linearization along the Predicted Trajectory (MPC-NPLPT) technique stands out. In MPC-NPLPT, the predicted output is a linear approximation calculated along some assumed future control sequence, and to enhance prediction accuracy and control quality, linearization is performed repeatedly. In this paper, the Adaptive Model Predictive Control - Nonlinear Prediction and Linearization along the Predicted Trajectory (AMPC-NPLPT) with the ANN model was used to enhance prediction precision.
\section{Hybrid Deep Learning Neural Network Controller}
\begin{figure}[H]
\centerline{\includegraphics[scale=0.37]{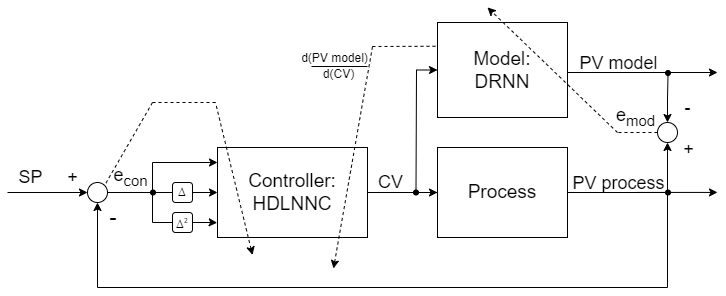}}
\caption{Scheme of HDLNNC.}
\label{fig:HDLNNC}
\end{figure}

Figure \ref{fig:HDLNNC} illustrates the HDLNNC scheme, which comprises two distinct blocks that respectively correspond to the model (Deep Recurrent Neural Network - DRNN) and the controller (Hybrid Deep Learning Neural Network Controller - HDLNNC).
\subsection{Controller}
The model block in HDLNNC scheme \cite{b2} is a deep learning model employing an Artificial Neural Network (ANN) comprising three layers in each of two parts. The first part consists of a self-organizing Kohonen map (SOM) with Hebbian learning, referred to as the Hybrid Deep Learning (HDL) layer. The second part comprises a Multi-Layer Feed-Forward Neural Network (MLFFNN), incorporating an adaptive learning factor for weight updates. The input signals to the ANN are defined as the error signal, the rate of change of the error signal, and the rate of change of the error signal at each sampling instance.
\par At the onset of each instance, the initial step involves computing the controller error, denoted by $e_{con}$. Subsequently, the weight update for the first two layers of the HDL component is determined as
\begin{equation}
W_{j}(k+1)=W_{j}(k) - h_{winner,j}(k)(X(k) - W_{j}(k)) \in R^{N_L} \label{eq 1}
\end{equation}
Equation \ref{eq 1} is utilized in the SOM learning process, whereby the weight values are updated via the winner-takes-most (WTM) approach. The update of weight values $W_j(k)$ is determined not only for the winning neuron, i.e., the one that exhibits weight values closest to the input values $X(k) \in R^{N_L}$ of the ANN in a selected metric but also for the current iteration denoted by $k$. Where $N_L$ denotes the number of neurons in the previous layer. Function $h_{winner,j}(k)$ is calculated as
\begin{equation}
h_{winner,j}(k)=l(k)e^{-\frac{r^{2}_{winner,j}(k)}{2\xi(k)^{2}}} \in R^{N_L} 
\label{eq 2}
\end{equation}
where:
\begin{equation}
l(k)=e^{-\frac{k}{K_L}} \in R\label{eq 3}
\end{equation}
\begin{equation}
\xi(k)=\xi_0\left({-\frac{\xi_f}{\xi_0}}\right)^{\frac{k}{K_L}} \in R\label{eq 4}
\end{equation}
where $r_{winner,j}$ denotes the distance between the $j^{th}$ neuron and the winning neuron, $K_{L}$ denotes the maximum number of learning samples and $l_0$,$\xi_0$, $\xi_f$  are coefficients selected from range $(0;1>$. 
\par The weights in the last layer, i.e. in the third layer in HDL are updated using the Hebbian learning procedure as follows
\begin{equation}
W_{j}(k+1)=W_{j}(k) + \gamma Y(k)X(k) - \delta Y(k)W(k) \in R^{N_L}\label{eq 5}
\end{equation}
Where $Y(k) \in R^{N_O}$ means the previous output from the ANN, $N_O$ the number of neurons in the layer, $\delta$ and $\gamma$ are coefficients from range (0,1). The part $\delta Y(k)W(k)$ is responsible for forgetting a learned pattern. 
\begin{equation}
\begin{array}{ll}
CV(k)=f(
f(\begin{bmatrix}
e_{con}(k) \\
\Delta e_{con}(k) \\
\Delta^{2} e_{con}(k) \\
\end{bmatrix}W^{I}_{HDL}(k))\\W^{II}_{HDL}(k))W^{III}_{HDL}(k) \in R
\end{array}
\label{new eq 6}
\end{equation}
\par Upon completion of the weight update procedure in the HDL component, the subsequent step involves computing the control variable, as specified in equation \eqref{new eq 6}. Herein, $W^{I}_{HDL}(k)$, $W^{II}_{HDL}(k)$, and $W^{III}_{HDL}(k)$ represent the weights of each layer at the current instant $k$, as per equations \eqref{eq 1} and \eqref{eq 5}. The activation function, denoted as $f()$, is chosen as the tangent hyperbolic function. Assuming that the Mean Squared Error is the minimizing function, the weights in the MLFFNN are modified as
\begin{equation}
\begin{array}{ll}
W(k+1)=W(k) + \eta(k) e_{con}(k) \\ \frac{\sigma PV_{process}(k)}{\sigma CV(k)}\frac{\sigma CV(k)}{\sigma W(k)} \in R^{N_L \times N_A}\label{eq 6}
\end{array}
\end{equation}
where
\begin{equation}
\eta(k)=\frac{\alpha}{\phi (1 + \frac{\beta}{\phi}(\frac{\sigma e_{con}(k)}{\sigma W(k)})^2)} \in R^{N_L \times N_A} \label{eq 7}
\end{equation}
Where $N_A$ denotes the number of neurons in the actual layer.
\par Equation \eqref{eq 7} is the result of the selection of the Lyapunov function as \eqref{eq 8} in which the value of $\eta(k)$ is determined by one of the three conditions present: $V(k+1) - V(k) \leqslant 0$ in the direct Lyapunov method and satisfies the other two $V(k) > 0, V(0) = 0$ as
\begin{equation}
V(k)=\frac{\alpha}{2} e_{con}(k)^2 + \frac{\beta}{2} \Delta e_{con}(k)^2 + \frac{\phi}{2} \Delta W(k)^2 \in R \label{eq 8}
\end{equation}
Where in Equation \eqref{eq 7}, \eqref{eq 8} $e_{con}$ means the error of control in current instant, ${\alpha}$, ${\beta}$, ${\phi}$ are coefficients selected between (0:1$>$, Jacobian of model can be estimated by
\begin{equation}
\frac{\sigma PV_{process}(k)}{\sigma CV(k)}\approx \frac{\sigma PV_{model}(k)}{\sigma CV(k)} \in R
\end{equation}
\subsection{Model}
The adaptive learning rate in the DRNN model \cite{b5} was calculated by 
\begin{equation}
\eta_{O}(k)=\frac{2}{N_{h}} \in R \label{eq 9}
\end{equation}
\begin{equation}
\eta_{D}(k)=\frac{2}{N_{h}(max||W^{0}(k)||)^{2}} \in R\label{eq 10}
\end{equation}
\begin{equation}
\eta_{I}(k)=\frac{2}{N_{h}(max||W^{0}(k)||)^{2}(max||I^{0}(k)||)^{2}} \in R\label{eq 11}
\end{equation}
Where $N_h$ denotes the number of neurons in only the hidden layer, $W^{O}(k)$ weights in the output layer, $I^{O}(k)$ weights in input to the neural network model, $\eta_{O}(k)$, $\eta_{D}(k)$, $\eta_{I}(k)$ adaptive learning  rate respectively for output, diagonal, input layer. In this case, the selected Lyapunov function is
\begin{equation}
V(k)=\frac{1}{2} e_{mod}(k)^2 \in R \label{eq 12}
\end{equation}
Table \ref{parameters} includes applied parameter values. Initial values for the weights of the controller and model, also the initial values of the diagonal layer in the DRNN model, were drawn from range (-0.1, 0.1). 
%

\section{Adaptive Model Predictive Control – Nonlinear Prediction and Linearization Along The Predicted Trajectory}
The MPC control algorithm offers several significant advantages, including its capacity to handle Multiple-Input Multiple-Output (MIMO) processes. Additionally, the algorithm is robust to process delays, and it delivers high-quality control performance even in the presence of such delays. Another advantage of the MPC algorithm is its ability to accommodate constraints on the input and output variables or state variables. This feature is particularly useful in practical control applications where constraints are commonly encountered and must be satisfied to ensure optimal performance and safety.
\par The basic idea of MPC algorithm \cite{b7} is to use the online dynamics of the model of the process to calculate predicted errors over the given horizon and minimize the cost function
\begin{equation} 
\begin{split}
J_{\text{CF}}(k)=\sum_{p=1}^{N} (y_{sp}(k+p|k) - y_{\sim}(k+p|k))^2  + \\
+\lambda \sum_{p=0}^{N_U-1} (\Delta u(k+p|k))^2 \in R \label{eq 13}
\end{split}
\end{equation}
where $y_{sp}(k+p|k)$ refers to the setpoint at step $k+p$, which is known at the current step $k$, for the process output, $y_{\sim}(k+p|k)$ denotes the predicted value at step $k+p$ for the output of the process model, known at the current step $k$, $\lambda$ is a user-defined weighting factor that is used to adjust the relative importance of the setpoint tracking and control effort optimization objectives, the prediction horizon is denoted by $N$, and $N_{u}$ is the control horizon, the quantity $\Delta u(k+p|k)$ denotes the change in the control signal over the prediction horizon.
\par The imposed constraints are represented by 
\begin{equation}
J_{\text{CF}}(k)=||Y_{sp}(k) - Y_{\sim}(k)||^{2} + ||\Delta U(k)||^{2}_{\Lambda} \in R \label{eq 14}
\end{equation}
only if
\begin{equation}
U_{min} \leq U(k) \leq U_{max}\label{eq 15}
\end{equation}
\begin{equation}
\Delta U_{min} \leq U(k) \leq \Delta U_{max}\label{eq 16}
\end{equation}
\begin{equation}
Y_{min} \leq Y_{\sim}(k) \leq Y_{max}\label{eq 17}
\end{equation}
where
\begin{equation}
Y_{sp}(k) = [y_{sp}(k) \dots y_{sp}(k+N)]^{T} \in R^{N}\label{eq 18}
\end{equation}
\begin{equation}
Y_{\sim}(k) = [y_{\sim}(k+1|k) \dots y_{\sim}(k+N|k)]^{T} \in R^{N}\label{eq 19}
\end{equation}
\begin{equation}
U_{min} = [u_{min} \dots u_{min}]^{T} \in R^{N_{u}}\label{eq 20}
\end{equation}
\begin{equation}
U_{max} = [u_{max} \dots u_{max}]^{T} \in R^{N_{u}}\label{eq 21}
\end{equation}
\begin{equation}
\Delta U_{max} = [\Delta u_{max} \dots \Delta u_{max}]^{T} \in R^{N_{u}}\label{eq 22}
\end{equation}
\begin{equation}
Y_{min} = [y_{min} \dots y_{min}]^{T} \in R^{N}\label{eq 23}
\end{equation}
\begin{equation}
Y_{max} = [y_{max} \dots y_{max}]^{T} \in R^{N}\label{eq 24}
\end{equation}
\begin{equation}
\Lambda = diag(\lambda \dots \lambda) \in R^{N_u \times N_u}\label{eq 25}
\end{equation}
where $u_{min}(k)$, $u_{max}(k)$ denotes the minimum/maximum value of the control variable, $y_{min}(k)$, $y_{max}(k)$ denotes the minimum/maximum of the value of output variable.
\begin{figure}[H]
\centerline{\includegraphics[scale=0.35]{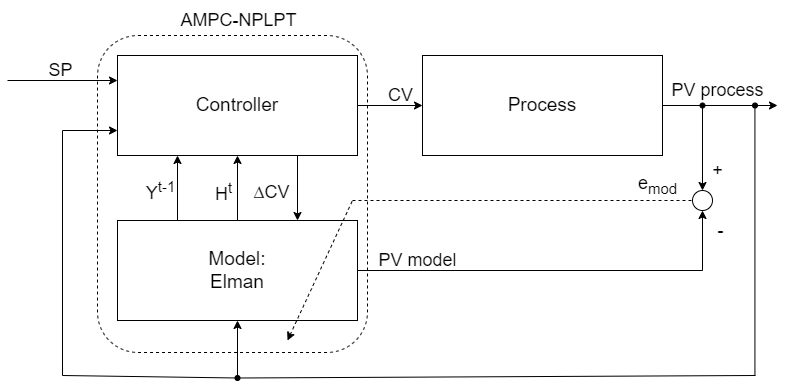}}
\caption{Scheme of AMPC-NPLPT.}
\label{Scheme of AMPC-NPLPT}
\end{figure}
\begin{table}[H]
\caption{Parameters HDLNNC and AMPC-NPLPT}
\begin{center} 
\begin{tabular}{|c|c|c|c|}
\hline
\textbf{Parameter} & \textbf{Value} & \textbf{Parameter} & \textbf{Value} \\
\hline
\multicolumn{2}{|c|}{ HDLNNC } & \multicolumn{2}{|c|}{ AMPC-NPLPT } \\
\hline
$\alpha$ & $9 \cdot 10^{-2}$  & $N$ & $15$ \\
\hline
$\beta$ & $9 \cdot 10^{-1}$ & $N_u$ & $3$\\
\hline
$\phi$ & $10^{-1}$ & $\lambda$ & $1$\\
\hline
$l_0$ & $5 \cdot 10^{-5}$ & $\gamma$ & $10^{-7}$ \\
\hline
$\xi_0$ & $7 \cdot 10^{-5}$ & $\delta$ & $10^{-15}$ \\
\hline
$\xi_f$ & $5 \cdot 10^{-5}$ & $t$ & $10$ \\
\hline
$\gamma$ & $10^{-4}$ & $u_{min}$ & $-1.5$ \\
\hline
$\delta$ & $10^{-6}$ & $u_{max}$ & $1.5$ \\
\hline
 & & $\Delta u_{max}$ & $3 \cdot 10^{-1}$ \\
 \hline
 & & $y_{min}$ & $-1.5$ \\
 \hline
 & & $y_{max}$ & $1.5$ \\
\hline
\end{tabular}
\label{parameters}
\end{center}
\end{table}
Regarding the MPC-NPLPT and AMPC-NPLPT control algorithms (as shown in Figure \ref{Scheme of AMPC-NPLPT}), an internal iteration technique was employed to enhance both the accuracy of the predicted process output and the control performance. This involved repeated model linearization and calculation of the future control increment. Specifically, the Taylor series expansion approximates the nonlinear output with a linear model.
\begin{equation}
Y_{\sim}^{i}(k) = Y_{\sim}^{i-1}(k) + H^{i}(k) (U^{i}(k) - U^{i-1}(k))\label{eq 26}
\end{equation}
where:
\begin{equation}
Y_{\sim}^{i-1}(k) = [y_{\sim}^{i-1}(k+1|k) \dots y_{\sim}^{i-1}(k+N|k)]^{T} \in R^{N}\label{eq 27}
\end{equation}
\begin{equation}
\begin{array}{ll}
H^{i}(k)& =  \frac{dY_{\sim}^{i-1}(k)}{dU^{i-1}(k)} = \\
& =\begin{bmatrix}
\frac{dy_{\sim}^{i-1}(k+1|k)}{du^{i-1}(k|k)} & \dots & \frac{dy_{\sim}^{i-1}(k+1|k)}{du^{i-1}(k+N_{u}-1|k)}\\
\vdots & \ddots &  \vdots \\ 
\frac{dy_{\sim}^{i-1}(k+N|k)}{du^{i-1}(k|k)} & \dots & \frac{dy_{\sim}^{i-1}(k+N|k)}{du^{i-1}(k+N_{u}-1|k)}
\end{bmatrix} \in R^{N \cdot N_{u}} 
\end{array}
\label{eq 28}
\end{equation}
\begin{equation}
U^{i}(k) = [u^{i} (k|k) \dots u^{i} (k+N|k)]^{T} \in R^{N_u}\label{eq 29}
\end{equation}
\begin{equation}
U^{i-1}(k) = [u^{i-1} (k|k) \dots u^{i-1} (k+N|k)]^{T} \in R^{N_u}\label{eq 30}
\end{equation}
where $i$ means the internal iteration of the MPC-NPLPT algorithm. Vector $U^{i}(k)$ is calculated for $i = 0$ (i.e. the first iteration) as
\begin{equation}
U^{0}(k) = U^{i-1}(k) = [u(k-1) \dots u(k-1)]^{T} \in R^{N_u}\label{eq 31}
\end{equation}
and for $i$ greater than zero is calculated as
\begin{equation}
U^{i}(k) = J \Delta U^{i}(k) + U(k-1) \in R^{N_u}\label{eq 32}
\end{equation}
where $J$ is all one lower triangular matrix. 
This study utilised an Artificial Neural Network (ANN) model with an Elman structure to represent the control process. When investigating the impact of process delays on the MPC-NPLPT and AMPC-NPLPT algorithms, a modified Elman structure was employed \cite{b9}, which delayed the input to account for the processing delay. The Armijo rule  \cite{b1} was applied to determine the optimal learning rate for the model, both before and during the control process. The parameter values used for the AMPC-NPLPT algorithm are presented in Table \ref{parameters}. Where $\gamma$, $\delta$ means values below which the internal iteration is interrupted due to either a small change in the control signal or sufficient control accuracy. Specifically, the initial weights and the initial values of the hidden layer output were randomized within the range of (-0.1, 0.1).
\section{Simulation Results}
The controllers HDLNNC, AMPC-NPLPT were used to control the following nonlinear process described as follows
\begin{equation}
x_{1}(k) = A_{1}x_{1}(k-1) + A_{2}x_{2}(k-1) \label{eq 33}
\end{equation}
\begin{equation}
x_{2}(k) = \frac{x_{1}(k-1)}{A_{3} + x_{1}(k-1)^{2} + x_{2}(k-1)^{2}} \label{eq 34}
\end{equation}
\begin{equation}
y(k) = x_{1}(k) \label{eq 35}
\end{equation}
\par The initial values of the constants $A_{1}$, $A_{2}$, $A_{3}$ were taken as: 0.2, 0.8, 1.1 and the values of states as $x_{1}(0) = 0$ and $x_{2}(0) = 0$. A sinusoidal signal with an amplitude of 1 and a frequency  $\frac{\pi}{4}$ was used as a reference signal from 0 to 100 seconds. In contrast, a rectangular signal with amplitudes between: -0.4, 0.4 was used and a cycle time of 4 seconds was used in 100-150 seconds, which was additionally filtered by an object with transfer function as $\frac{1}{0.025 \cdot s + 1}$. In addition, in the $100^{th}$ second the values of the constants: $A_{1}$, $A_{2}$, $A_{3}$ were changed to: -0.2, 1.4, -15.
\par An assessment of the control accuracy of the two controllers and a comparison between them was undertaken based on following Integral Control Quality Indices (ICQI) 
\begin{equation}
I_{IAE} = \int_{0}^{\infty} |e(t)| dt\label{eq 36}
\end{equation}
\begin{equation}
I_{ISE} = \int_{0}^{\infty} e(t)^{2} dt\label{eq 37}
\end{equation}
\begin{equation}
I_{ITAE} = \int_{0}^{\infty} t|e(t)| dt\label{eq 38}
\end{equation}
\par Before initiating control of the process, the Elman model used in the AMPC-NPLPT controller was trained by a sinusoidal signal with amplitudes of 0.8, 0.6, and 0.5, respectively, and a frequency of $\frac{\pi}{4}$ for each of these signals. The learning signal was updated after every 8 seconds, corresponding to a full period of each sinusoidal signal. The training was stopped once the target function's mean squared error (MSE) value fell below $10^{-15}$, achieved in approximately the $22^{nd}$ second. This procedure was intended to initialize the weights of the Elman model, which would be further adapted during the control process. The simulation was conducted in Simulink with a calculation step of 1 ms for the nonlinear process without delay.
\begin{table}[htb]
\caption{ICQI for the process without delay (where SIN - sinusoidal reference signal, SQR - square reference signal)}
\begin{center} 
\begin{tabular}{|l|r|r|r|r|}
\hline
ICQI $\downarrow$ & \multicolumn{2}{|c|}{ SIN }& \multicolumn{2}{|c|}{ SQR }\\
\hline
&  HDLNNC & AMPC &  HDLNNC & AMPC \\
\hline
Time (s) $\rightarrow$ & \multicolumn{2}{|c|}{0:8}& \multicolumn{2}{|c|}{ 100:104 }\\
\hline
$I_{IAE}$ & $4.32 \cdot 10^{-2}$ & $7.53 \cdot 10^{-2}$ & $3.03 \cdot 10^{-3}$ & $2.63 \cdot 10^{-3}$ \\
$I_{ISE}$ & $4.08 \cdot 10^{-4}$ & $4.45 \cdot 10^{-2}$ & $7.30 \cdot 10^{-5}$ & $5.93 \cdot 10^{-5}$ \\
$I_{ITAE}$ & $1.58\cdot 10^{-1}$ & $2.86 \cdot 10^{-2}$ & $3.09 \cdot 10^{-1}$  & $2.68 \cdot 10^{-1}$ \\
\hline
Time (s) $\rightarrow$ & \multicolumn{2}{|c|}{8:16}& \multicolumn{2}{|c|}{104:108}\\
\hline
$I_{IAE}$ & $3.34 \cdot 10^{-2}$ & $5.56 \cdot 10^{-3}$& $2.87\cdot10^{-3}$ & $2.54\cdot10^{-3}$ \\
$I_{ISE}$ & $2.05 \cdot 10^{-4}$ & $1.21 \cdot 10^{-5}$ & $6.48\cdot10^{-5}$ & $5.72\cdot10^{-5}$\\
$I_{ITAE}$ & $4.02\cdot 10^{-1}$ & $6.72 \cdot 10^{-2}$ & $3.04\cdot10^{-1}$  & $2.68\cdot10^{-1}$\\
\hline
Time (s) $\rightarrow$ & \multicolumn{2}{|c|}{32:40}& \multicolumn{2}{|c|}{116:120}\\
\hline
$I_{IAE}$ & $2.54 \cdot 10^{-2}$ & $5.52 \cdot 10^{-3}$ & $2.61\cdot10^{-3}$ & $1.46\cdot10^{-3}$ \\
$I_{ISE}$ & $1.19 \cdot 10^{-4}$ & $1.16 \cdot 10^{-5}$ & $5.57\cdot10^{-5}$ & $2.04\cdot10^{-5}$  \\
$I_{ITAE}$ & $9.17 \cdot 10^{-1}$ & $1.99 \cdot 10^{-1}$  & $3.08\cdot10^{-1}$ & $1.71\cdot10^{-1}$ \\
\hline
Time (s) $\rightarrow$ & \multicolumn{2}{|c|}{88:96}& \multicolumn{2}{|c|}{144:148}\\
\hline
$I_{IAE}$ & $1.62 \cdot 10^{-2}$ & $5.41 \cdot 10^{-3}$ & $2.32\cdot10^{-3}$ & $6.66\cdot10^{-4}$ \\
$I_{ISE}$ & $4.91 \cdot 10^{-5}$ & $1.07 \cdot 10^{-5}$ & $5.07\cdot10^{-5}$ & $8.16\cdot10^{-6}$ \\
$I_{ITAE}$ & $1.50$ & $4.98 \cdot 10^{-1}$ & $3.39\cdot10^{-1}$ & $9.73\cdot10^{-2}$\\
\hline
\end{tabular}
\end{center}
\label{tab:no_delay}
\end{table}
 \begin{figure}[H]
\centerline{\includegraphics[scale=0.175]{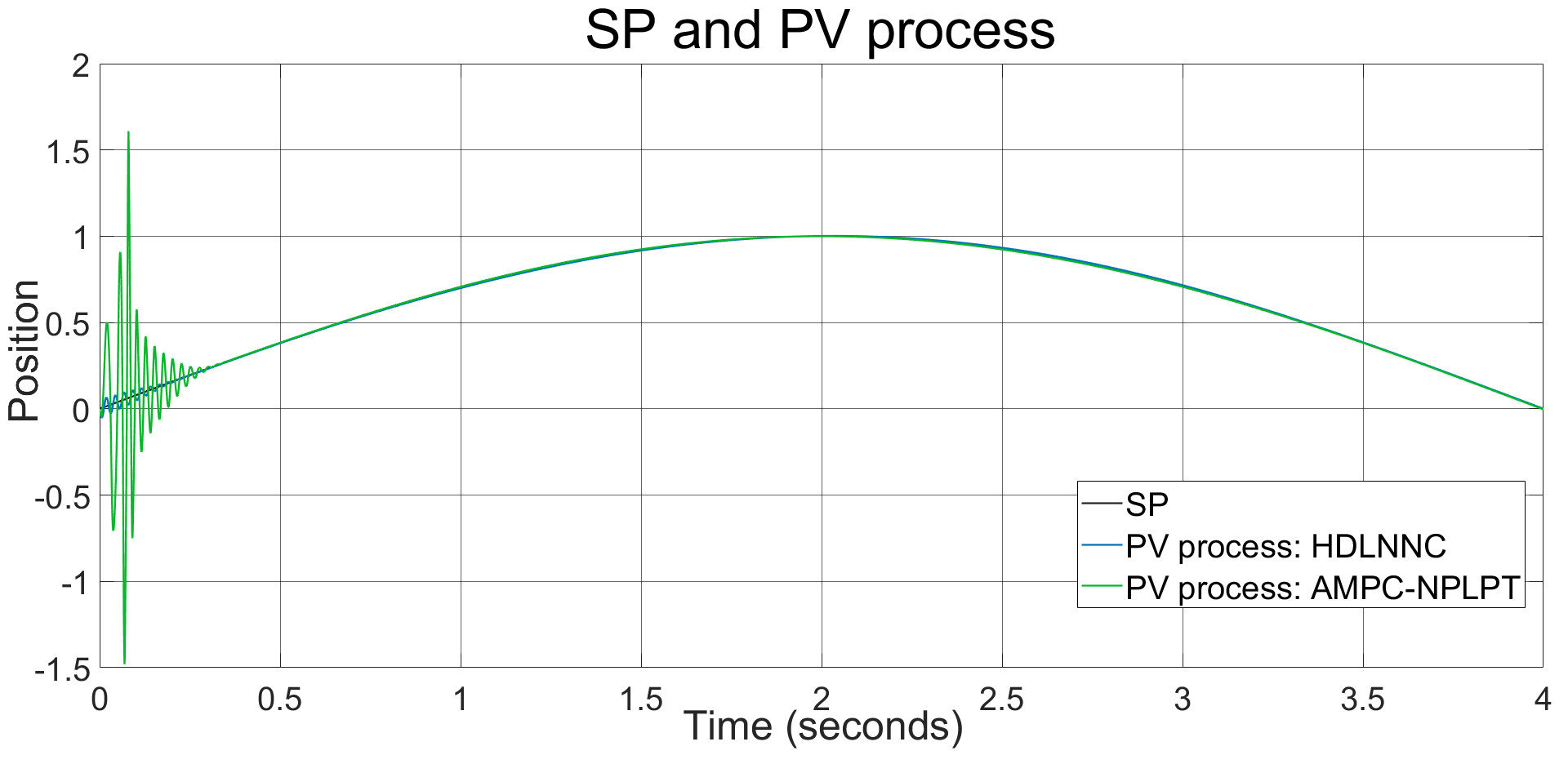}}
 \caption{Reference signal and process output for HDLNNC and AMPC-NPLPT in time from 0 to 4 seconds.}
 \label{fig:f3}
 \end{figure}
\par In the AMPC-NPLPT controller, the only hidden layer had 5 neurons. The prediction horizon was set to 15 instants, and the control horizon was set to 3 instants. A weighting factor of 1 was used for $\lambda$. The number of internal iterations was 10.
\par Table \ref{tab:no_delay} presents the quality indicators for the control actions taken by both controllers from 0 to 100 seconds. In almost every case, the AMPC-NPLPT controller produced smaller ICQI values, indicating more accurate control.
\par Due to changes in the object's dynamics, which were calculated from changes in the DRNN's weights, the HDLNNC controller resulted in a smoother start of control than the AMPC-NPLPT controller. The AMPC-NPLPT controller adjusted its weights significantly in the initial phase based on the output of the nonlinear object (Fig. \ref{fig:f3}). After about 0.3 seconds, the AMPC-NPLPT controller produced more accurate tracking of the reference signal than the HDLNNC controller. However, because the model inaccurately represented the object's behavior up to 0.3 seconds, it caused the output to exceed the acceptable limits of the $y_{max}$ value.
\par
\begin{figure}[htb]
\centerline{\includegraphics[scale=0.173]{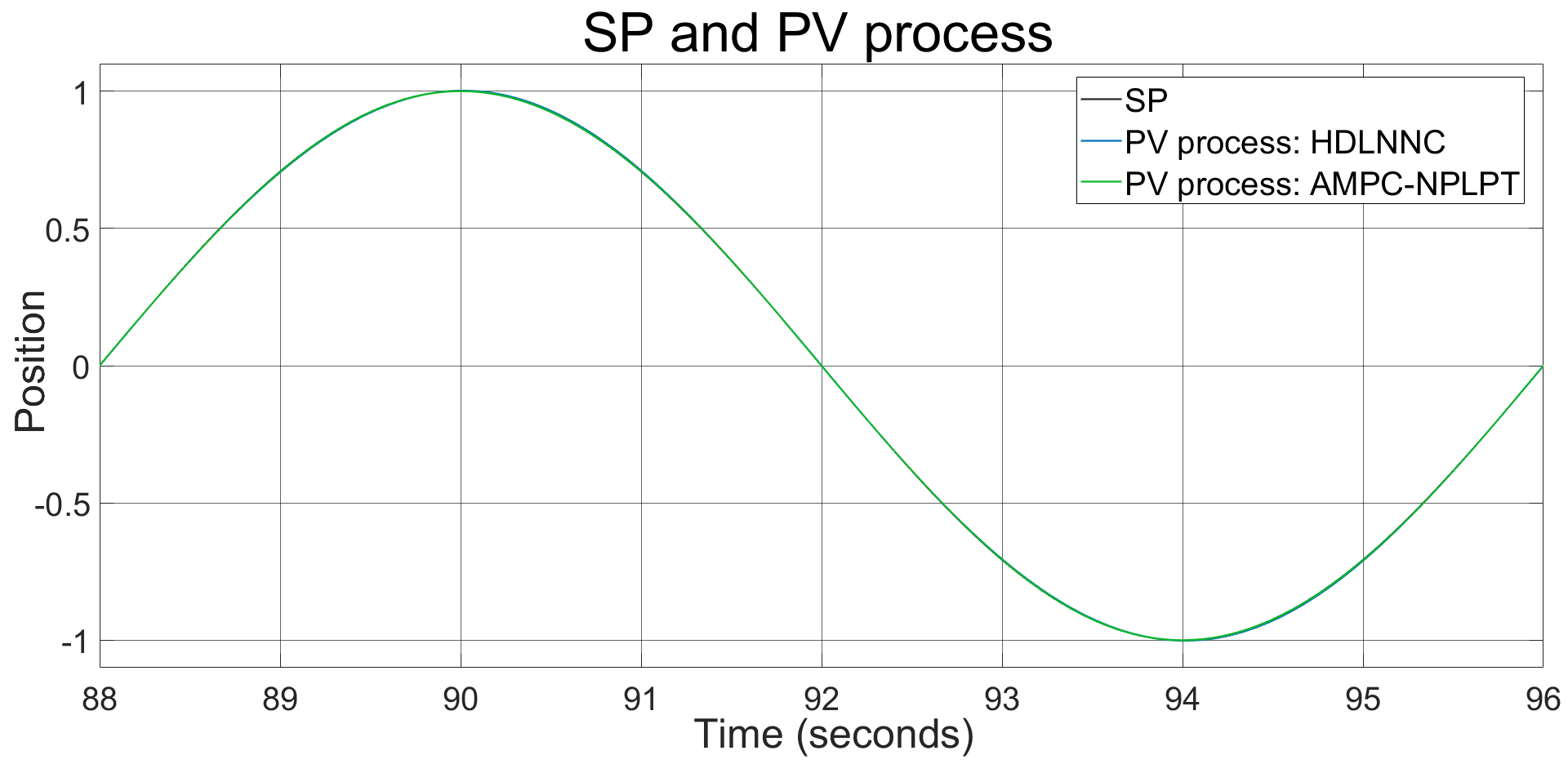}}
\caption{Reference signal and process output for HDLNNC and AMPC-NPLPT from 88 to 96 seconds.}
\label{fig:f4}
\end{figure}

\begin{figure}[htb]
\centerline{\includegraphics[scale=0.175]{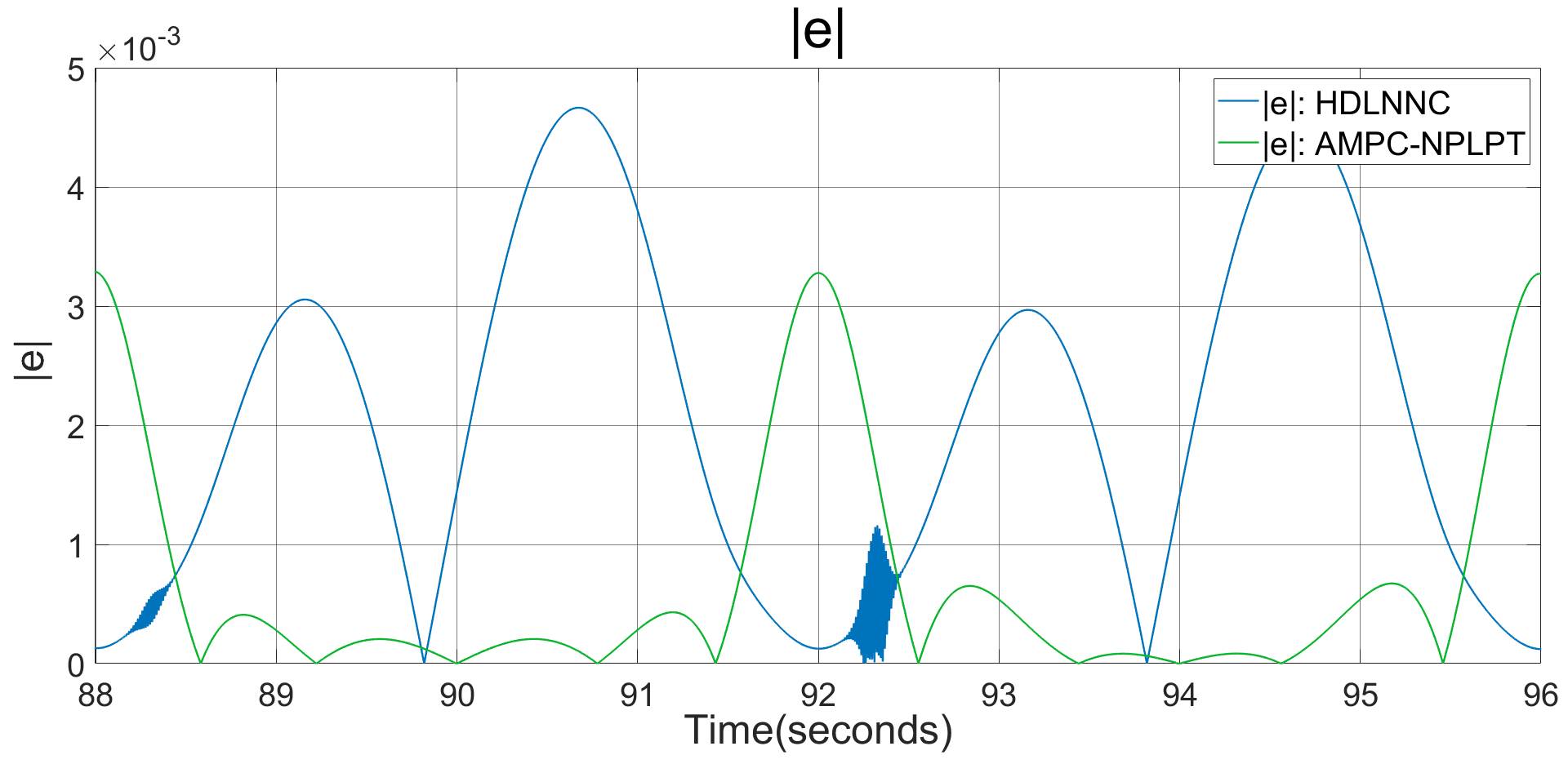}}
\caption{Absolute error for HDLNNC and AMPC-NPLPT from 88 to 96 seconds - sinusoidal refernece signal.}
\label{fig:fesin}
\end{figure}

\begin{figure}[htb]
\centerline{\includegraphics[scale=0.175]{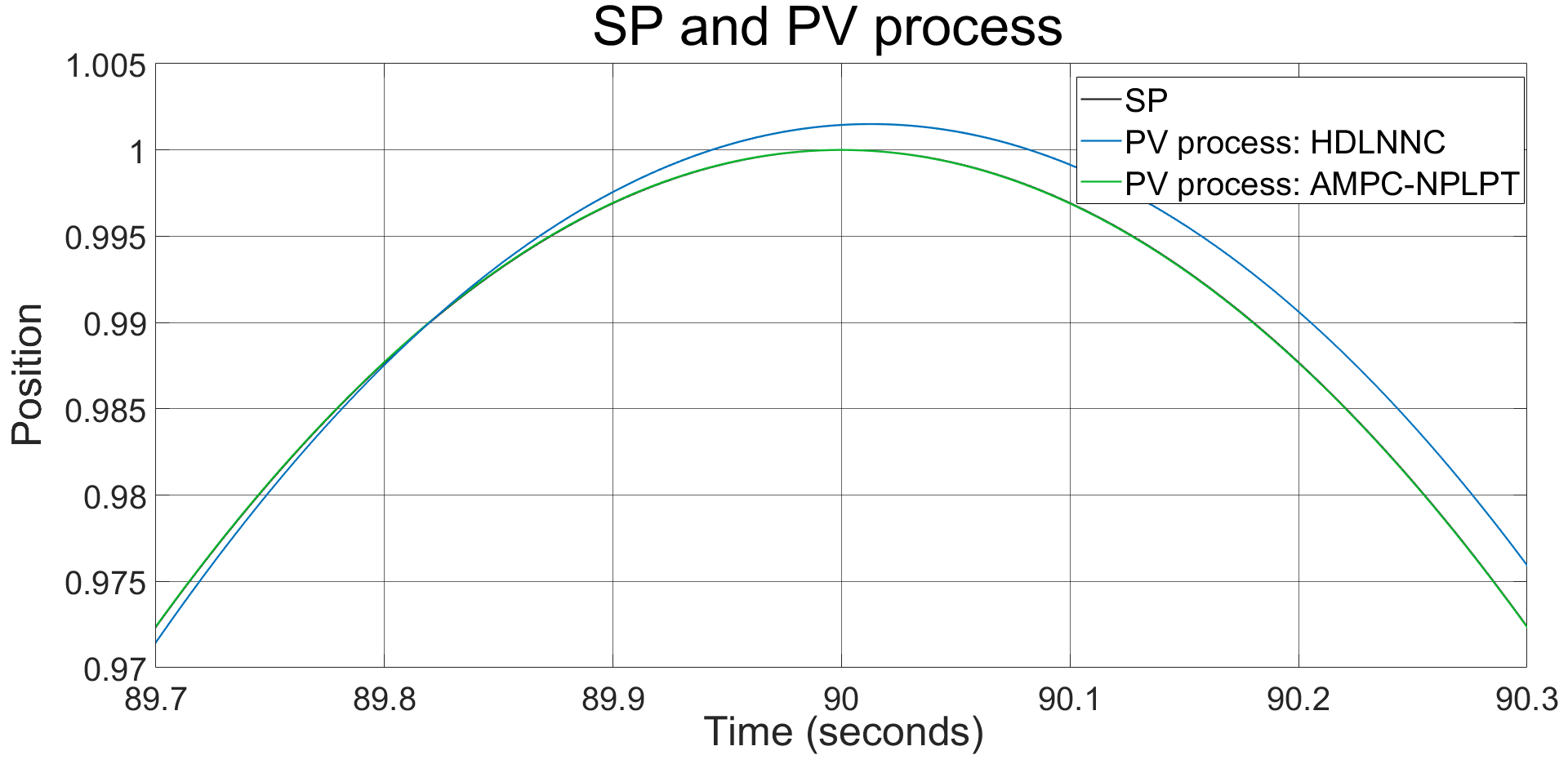}}
\caption{Reference signal and process output for HDLNNC and AMPC-NPLPT from 89.7 to 90.3 seconds.}
\label{fig:f5}
\end{figure}

 Figure \ref{fig:f4} compares the position of the nonlinear object using the AMPC-NPLPT controller and the HDLNNC controller from 88 to 96 seconds. Figure \ref{fig:fesin} presents a comparison of the absolute error values at the same time. Figure \ref{fig:f5} presents a close-up between 89.7 and 90.3 seconds, showing that the AMPC-NPLPT controller provided more precise control. 
\par Despite the change in the coefficients describing the behaviour of the nonlinear object as well as the reference signal, which occurred in the $100^{th}$ second, an accurate control was obtained using both controllers. In this case, smaller values for all ICQI values were obtained using the controller: AMPC-NPLPT (Table \ref{tab:no_delay}). 
\begin{figure}[htb]
\centerline{\includegraphics[scale=0.175]{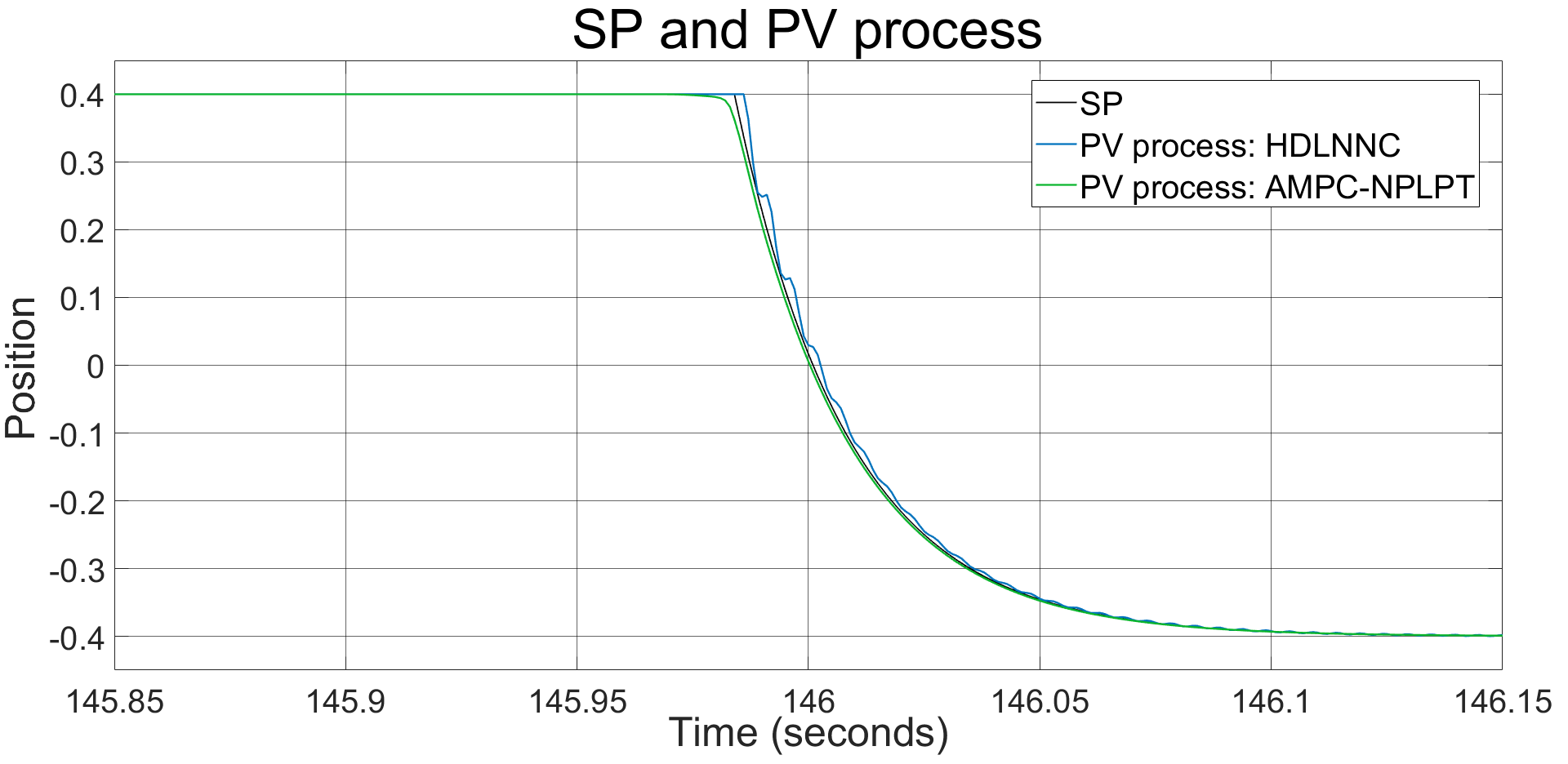}}
\caption{Reference signal and process output for HDLNNC and AMPC-NPLPT in 
time from 145.85 to 148.15 seconds  - square reference signal.}
\label{fig:f7}
\end{figure}
\begin{figure}[htb]
\centerline{\includegraphics[scale=0.175]{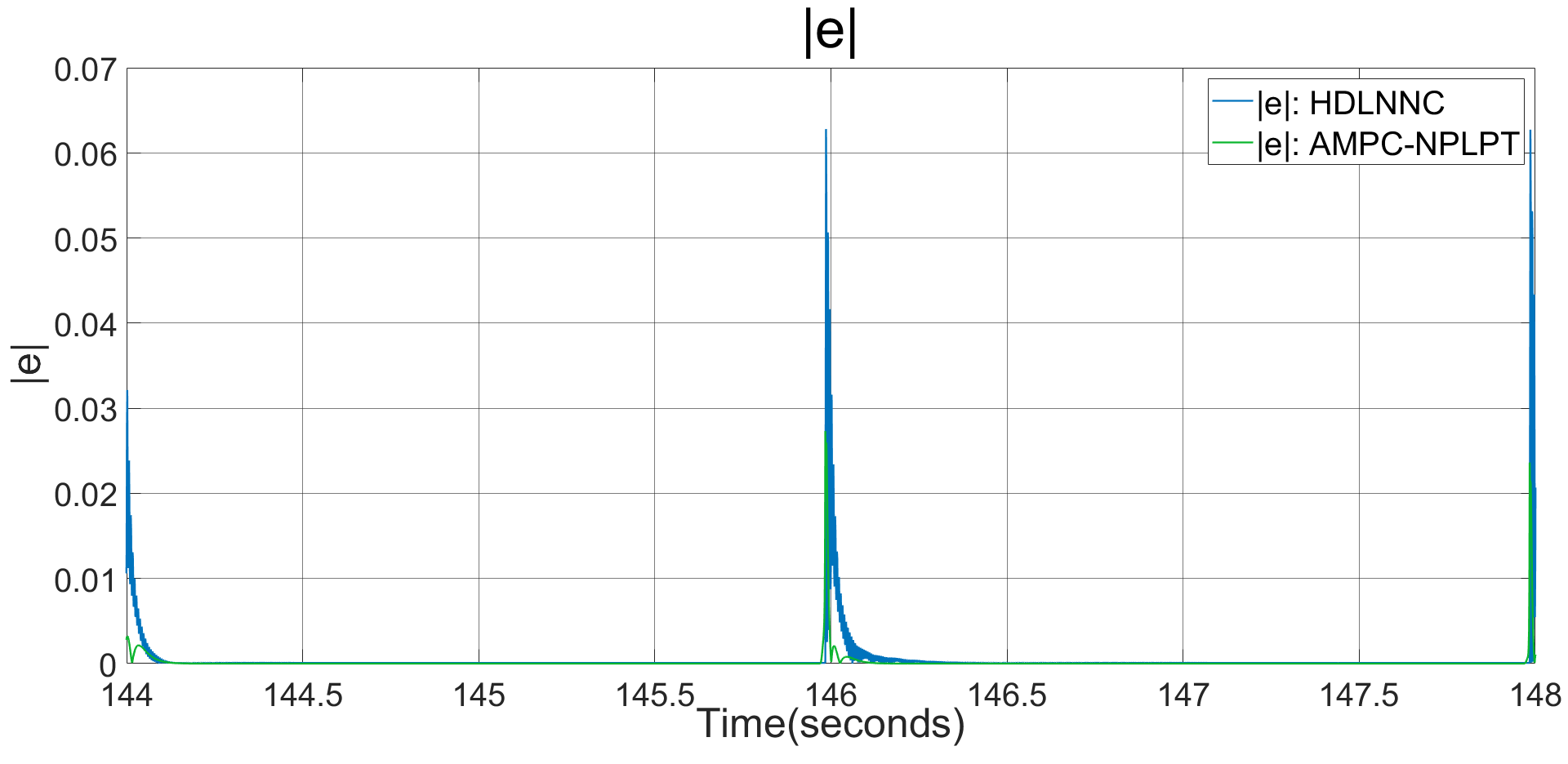}}
\caption{Absolute error for HDLNNC and AMPC-NPLPT from 144 to 148 seconds - square reference signal.}
\label{fig:fesq}
\end{figure}
\begin{table}[htb]
\caption{ICQI for the process with delay (where SIN - sinusoidal reference signal, SQR - square reference signal)}
\begin{center} 
\begin{tabular}{|l|r|r|r|r|}
\hline
ICQI $\downarrow$ & \multicolumn{2}{|c|}{ SIN }& \multicolumn{2}{|c|}{ SQR }\\
\hline
&  HDLNNC & AMPC &  HDLNNC & AMPC \\
\hline
Time (s) $\rightarrow$ & \multicolumn{2}{|c|}{0:8}& \multicolumn{2}{|c|}{ 200:204 }\\
\hline
$I_{IAE}$ & $4.54$ & $1.31$ & $2.41\cdot10^{1}$ & $6.32\cdot10^{-1}$\\
$I_{ISE}$ & $3.59$ & $3.92\cdot10^{-1}$ & $1.46\cdot10^{2}$ & $1.16\cdot10^{-1}$  \\
$I_{ITAE}$ & $9.17\cdot10^{2}$  & $2.10\cdot10^{2}$ & $2.44\cdot10^{5}$  & $6.83\cdot10^{3}$\\
\hline
Time (s) $\rightarrow$ & \multicolumn{2}{|c|}{8:16}& \multicolumn{2}{|c|}{204:208}\\
\hline
$I_{IAE}$ & $5.40$ & $1.44$ & $2.46\cdot10^{1}$ & $5.75\cdot10^{-1}$ \\
$I_{ISE}$ & $5.31 $ & $3.85\cdot10^{-1}$ & $1.52\cdot10^{2}$ & $9.58\cdot10^{-2}$ \\
$I_{ITAE}$ & $2.92\cdot10^{3}$  & $8.68\cdot10^{2}$ & $2.53\cdot10^{5}$  & $5.93\cdot10^{3}$ \\
\hline
Time (s) $\rightarrow$ & \multicolumn{2}{|c|}{32:40}& \multicolumn{2}{|c|}{216:220}\\
\hline
$I_{IAE}$ & $3.68\cdot10^{1}$ & $2.61$ & $2.46\cdot10^{1}$ & $3.39\cdot10^{-1}$ \\
$I_{ISE}$ & $1.79\cdot10^{2}$ & $1.15$ & $1.51\cdot10^{2}$ & $3.62\cdot10^{-2}$ \\
$I_{ITAE}$ & $6.66\cdot10^{4}$ & $4.66\cdot10^{3}$ & $2.68\cdot10^{5}$ & $3.69\cdot10^{3}$ \\
\hline
Time (s) $\rightarrow$ & \multicolumn{2}{|c|}{88:96}& \multicolumn{2}{|c|}{244:248}\\
\hline
$I_{IAE}$ & $3.88\cdot10^{1}$ & $2.16$ & $2.48\cdot10^{1}$ & $4.28\cdot10^{-1}$ \\
$I_{ISE}$ & $1.94\cdot10^{2}$ & $7.41\cdot10^{-1}$ & $1.54\cdot10^{2}$ & $5.83\cdot10^{-2}$\\
$I_{ITAE}$ & $1.78\cdot10^{5}$ & $9.92\cdot10^{3}$ & $3.05\cdot10^{5}$ & $5.27\cdot10^{3}$ \\
\hline
Time (s) $\rightarrow$ & \multicolumn{2}{|c|}{136:144}& \multicolumn{2}{|c|}{320:324}\\
\hline
$I_{IAE}$ & $3.89\cdot10^{1}$ & $1.17$ & $2.50\cdot10^{1}$ & $2.96\cdot10^{-1}$\\
$I_{ISE}$ & $1.95\cdot10^{2}$ & $2.57\cdot10^{-1}$ & $1.56\cdot10^{2}$ & $3.03\cdot10^{-2}$\\
$I_{ITAE}$ & $2.73\cdot10^{5}$ & $8.22\cdot10^{3}$ & $4.02\cdot10^{5}$ & $4.77\cdot10^{3}$ \\
\hline
Time (s) $\rightarrow$ & \multicolumn{2}{|c|}{184:192}& \multicolumn{2}{|c|}{396:400}\\
\hline
$I_{IAE}$ & $3.89\cdot10^{1}$ & $8.93\cdot10^{-1}$ & $2.50\cdot10^{1}$ & $2.47\cdot10^{-1}$ \\
$I_{ISE}$ & $1.95\cdot10^{2}$ & $1.44\cdot10^{-1}$ & $1.56\cdot10^{2}$ & $2.44\cdot10^{-2}$   \\
$I_{ITAE}$  & $3.66\cdot10^{5}$ & $8.34\cdot10^{3}$ & $4.97\cdot10^{5}$ & $4.96\cdot10^{3}$ \\
\hline
\end{tabular}
\end{center}
\label{tab:delay}
\end{table}
Figure \ref{fig:fesq} shows a comparison of the absolute error values from 144 to 148 seconds. Figure \ref{fig:f7} depicts a portion of the control cycle at the end of the control time, which shows that the AMPC-NPLPT predictive controller begins to decrease the output signal of the nonlinear object at approximately 145.98 seconds.
\par Table \ref{tab:delay} presents the results of applying both controllers to a delayed nonlinear object. In this case, the output value of the nonlinear object was delayed by 10 steps, resulting in a delay of 0.5 seconds due to an increased sampling time from 1 ms to 50 ms. This change in sampling time aimed to increase the resultant time delay, reducing computation while ensuring accurate approximation of both sinusoidal and rectangular signals. The sinusoidal signal was extended to 200 seconds, followed by a rectangular signal used for up to 400 seconds. A linearly increasing signal was used to minimize the sudden change in the square reference signal.
\par At the 200th second, the controller constants $A_1$, $A_2$, and $A_3$ were changed from 0.2, 0.8, 1.1 to -0.2, 1.4, -15, respectively, as in the case of no delay. The number of neurons in the HDLNNC controller's hidden layers was increased from 10 to 15 and from 5 to 8 in the first and second hidden layers, respectively, while the number of neurons in the DRNN model was increased to 15. This treatment aimed to improve the approximation of the object's dynamics and increase the number of weights for more accurate control.
\par For the AMPC-NPLPT controller, the number of hidden layer neurons remained at 5, but the prediction horizon was increased to 30 and the control horizon to 5. The $\lambda$ coefficient was changed to 0.8, the number of internal iterations increased to 20, and the value of the $\Delta u$ parameter was reduced to $3.5 \cdot 10^{-2}$.
\par The modification of the AMPC-NPLPT controller's input, which involved delaying it by as many steps as the output of a nonlinear object in an artificial neural network with an Elman structure, enabled the control of the process. However, the HDLNNC controller, despite an increase in the number of neurons in the hidden layers, was unable to follow the sinusoidal reference signal from time 0 to 200 seconds. To compensate for the delay and obtain the smallest possible ICQI, the CV signal was changed between the maximum values of -5 and 5. This course of the CV caused significant changes in the object position values.
\begin{figure}[htb]
\centerline{\includegraphics[scale=0.175]{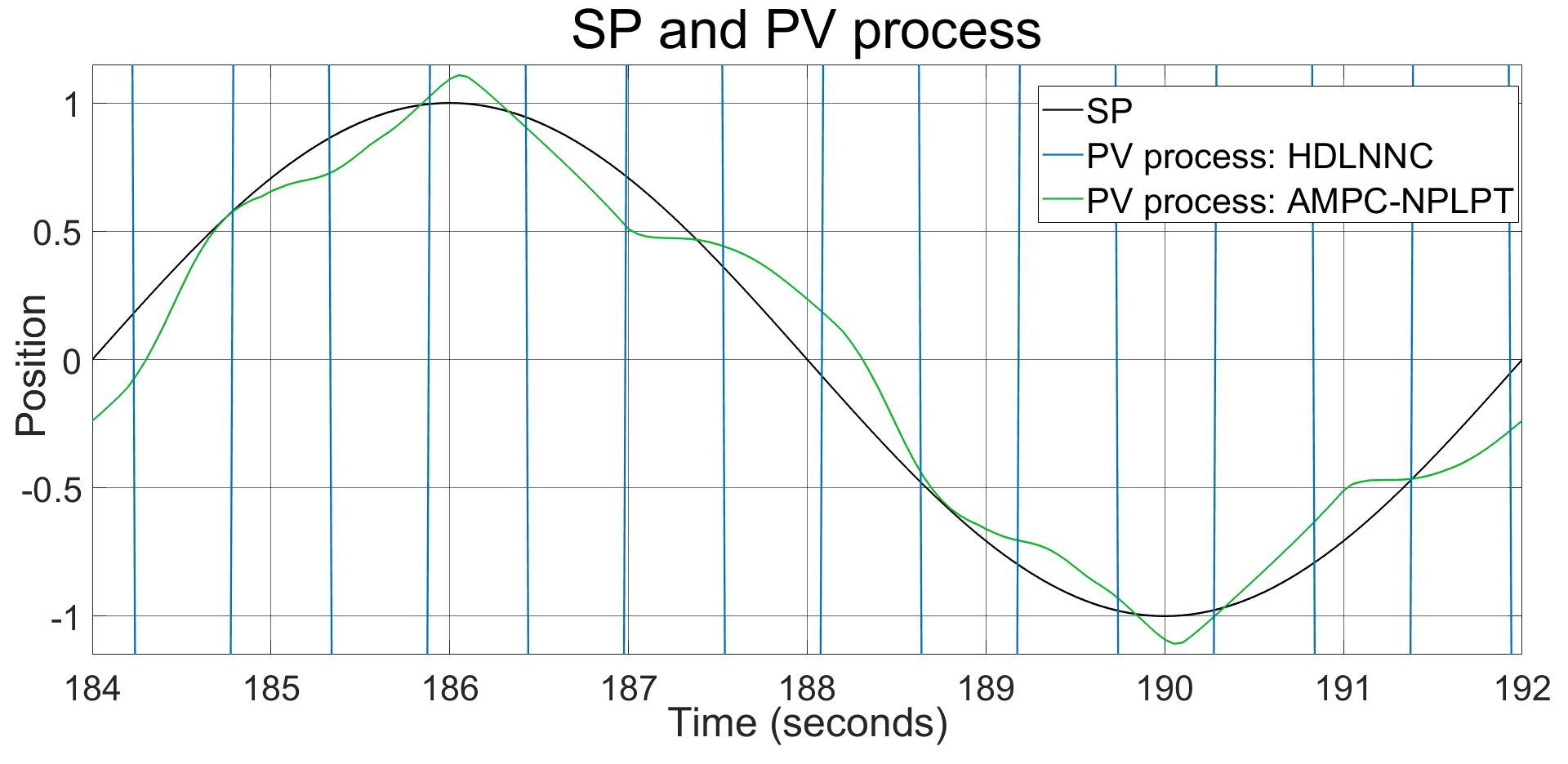}}
\caption{Reference signal and process output for HDLNNC and AMPC-NPLPT in 
time from 184 to 192 seconds.}
\label{fig:f8}
\end{figure}
%
%

%
%
Our study compared the AMPC-NPLPT and HDLNNC controllers' performance for a delayed nonlinear object. We found that the AMPC-NPLPT controller adjusts the control signal's course in a way that reduces the Integrated Control Quality Index (ICQI) from cycle five without exceeding the maximum control signal values, which in this case were -1 to 1 (as seen in Fig. \ref{fig:f8}). Moreover, when we used a modified rectangular signal as the reference signal with the predictive algorithm, the AMPC-NPLPT controller achieved more accurate control than the HDLNNC controller, as evidenced by the smaller ICQI values in Table \ref{tab:delay}.
In contrast, the HDLNNC controller aimed to reduce the error by changing the control signal values from -5 to 5, as it did for the sinusoidal signal. However, we observed that the total absolute error between the nonlinear object output and the DRNN model was 55.3831, indicating an average error per instant of $6.9228\cdot10^{-3}$.
\section{Conclusion}
The HDLNNC controller as well as the AMPC-NPLPT controller were successfully used to control the nonlinear object in the absence of delay. Using the AMPC-NPLPT controller, smaller ICQI values were obtained in most cases for the object in the absence of delay as well as with delay, even though it uses a single neural network and in the case of the HDLNNC controller the solution is more complex. Despite the increase in the number of neurons in both hidden layers, the HDLNNC controller was not able to control the selected nonlinear object with delay. Further research will be aimed at testing the performance of these controllers to other nonlinear objects even with delay as well as testing the absence of the previously learned AMPC-NPLPT controller. 
%

\end{document}